\newcommand{\nn}{\nonumber\\} 
\newcommand{\f}[1]{\mbox{\boldmath$#1$}}
\newcommand{\fk}[1]{\mbox{\boldmath$\scriptstyle#1$}}
\newcommand{\bea}{\begin{eqnarray}}
\newcommand{\ea}{\end{eqnarray}}
\newcommand{\ord}{{\cal O}}

\documentstyle[aps,prl,twocolumn,amstex]{revtex}

\begin{document}

\wideabs{
\title{Large-scale non-locality in ``doubly special relativity'' \\
with an energy-dependent speed of light}
%
\author{R.~Sch\"utzhold and W.~G.~Unruh}
\address{
Canadian Institute for Advanced Research Cosmology and Gravity Program,\\ 
Department of Physics and Astronomy, University of British Columbia,
Vancouver, British Columbia, Canada V6T 1Z1,\\
email: {\tt schuetz@@physics.ubc.ca; unruh@@physics.ubc.ca}}
\maketitle
\begin{abstract}
There are two major alternatives for violating the (usual) Lorentz invariance 
at large (Planckian) energies or momenta -- either not all inertial frames 
(in the Planck regime) are equivalent 
(e.g., there is an effectively preferred frame)
or the transformations from one frame to another are (non-linearly) deformed
(``doubly special relativity'').
We demonstrate that the natural (and reasonable) assumption of an 
energy-dependent speed of light in the latter method goes along with 
violations of locality/separability (and even translational invariance) 
on macroscopic scales.
\end{abstract}
}

\section{Introduction}
%
The observation that there is no invariant energy (and length, etc.) scale in 
special relativity on the one hand and the expected physical significance of 
the Planck scale on the other hand has motivated (see, e.g., 
\cite{Amelino-DSR1,Amelino-DSR2,Amelino-DSR3,Amelino-DSR4,Amelino-DSR5}
and \cite{Smolin-PRL,Smolin-PRD}) the suggestion that the (usual) Lorentz 
invariance might be broken at large (Planckian) energies.
It has also been suggested (see, e.g., 
\cite{Amelino-DSR1,Amelino-DSR2,Amelino-DSR3,Amelino-DSR4,Amelino-DSR5},
\cite{Smolin-PRL,Smolin-PRD}, and \cite{TeV1,TeV2,TeV3}) that several 
yet unexplained observations -- such as the indications for ultra-high energy 
cosmic rays (UHECR) with energies above the Greisen-Zatsepin-Kuzmin (GZK)
cut-off \cite{GKZ1,GKZ2} of order $\ord(10^{19}\,{\rm eV})$ induced by the 
interaction with the cosmic microwave background -- could be interpreted as 
empirical evidence for deviations from the Lorentz invariance at high energies.
Furthermore, variable speed-of-light (VSL) cosmologies 
(see, e.g., \cite{VSL1,VSL2}),
which also require breaking the local Poincar\'e symmetry, have been considered
as alternative solutions of the cosmological problems which lead to the idea 
of inflation. 
In particular in the first case, one needs to explain the apparent large gap 
between the energy range of those phenomena $\ord(10^{19}\,{\rm eV})$
and the Planck scale $\ord(10^{28}\,{\rm eV})$ -- but we are not going to 
discuss these phenomenological issues here.
(For phenomenological constraints on Lorentz violation see, e.g., 
\cite{Energy1,Energy2} and references therein, cf.~\cite{Kostelecky}).

There are two major alternatives for breaking the Lorentz invariance:
either not all inertial frames (in the Planck regime) are equivalent 
(e.g., there is an effectively preferred frame) or the transformations 
from one frame to another are different (deformed).
In the following we shall consider the second possibility in more detail and
discuss some consequences which arise thereof.
For the sake of simplicity (and since the masses of all known ``elementary'' 
particles are small compared to Planck scale) we shall consider massless 
particles, such as photons, only.
In addition, we shall work in 1+1 dimensions (unless otherwise noted).

The main idea of the ``doubly special relativity'' 
(DSR, see, e.g., 
\cite{Amelino-DSR1,Amelino-DSR2,Amelino-DSR3,Amelino-DSR4,Amelino-DSR5},
\cite{Smolin-PRL,Smolin-PRD}, and \cite{Visser}) 
is to replace the usual linear Lorentz transformation ${\cal L}$ by the 
following non-linear representation $F \circ {\cal L} \circ F^{-1}$, i.e., 
\bea
\label{non-lin}
\left(\begin{array}{c}  E \\ p \end{array}\right)
&\to&
\left(\begin{array}{c}  E' \\ p' \end{array}\right)
=
F \circ {\cal L} \circ F^{-1}
\left(\begin{array}{c}  E \\ p \end{array}\right)
\nn
&=&
F\left[
\frac{\left(\begin{array}{cc} 1 & v \\ v/c^2 & 1 \end{array}\right)}
{\sqrt{1-v^2/c^2}}\,
F^{-1}\left(\begin{array}{c}  E \\ p \end{array}\right)
\right]
\,,
\ea
with some non-linear function $F\,:\,{\mathbb R}^2\,\to\,{\mathbb R}^2$
\bea
\label{function}
\left(\begin{array}{c}  E \\ p \end{array}\right)
=
F
\left(\begin{array}{c} {\mathfrak E} \\ {\mathfrak p} \end{array}\right)
\,,
\ea
which reduces to the identity for small energies
\bea
\label{low-energy}
F\left(\begin{array}{c} 
E \ll M_{\rm Planck}\,c^2 \\ p \ll M_{\rm Planck}\,c
\end{array}\right)=
\left(\begin{array}{c}  E \\ p \end{array}\right)
\,.
\ea
Note that the group structure of the deformed transformations in 
Eq.~(\ref{non-lin}) is the same as that of the ordinary Lorentz group.
This appears quite reasonable as the only suitable six-parameter extension 
(cf.~\cite{Smolin-PRL,Smolin-PRD,Visser}) of the group $SO(3)$ of spatial 
rotations (which we want to retain) seems to be the Lorentz group itself -- 
especially since we want to reproduce the usual Lorentz transformations 
at small energies.
It should also be mentioned here that this approach relies on the particle 
picture -- there is no (unique and well-defined) field-theoretic formulation 
at this stage.

\section{Field-theoretic example}
%
Unfortunately, there is no unique prescription (so far) for translating the
behavior in momentum space $(E,p)$ into position space $(t,x)$, which is
required for formulating a corresponding field theory.
There is not even consistency in the literature regarding the velocity of
propagation of Planckian particles: 
In Ref.~\cite{c=1}, it is argued that the speed of light does not depend on 
the energy (i.e., that all massless particles have the same velocity $c$) 
in all DSR theories.
Ref.~\cite{c=dE/dp}, on the other hand, arrives at the (natural) result that 
the propagation speed is given by the group velocity $v_{\rm g}=dE/dp$.
In Ref.~\cite{Smolin-PRL}, however, the phase velocity $v_{\rm p}=E/p$ 
is used instead (in some limit).
This depends on whether (and how) one modifies the commutators such as 
$[x,p]=i\hbar$ and hence the identifications 
$ip\leftrightarrow\hbar\partial/\partial x$, 
etc., or not (see, e.g., \cite{Non-commu1,Non-commu2}).

However, let us consider one possible example for a field-theoretic
formulation motivated by an analogy to condensed matter systems.
The propagation of sound waves is governed by a dispersion relation which is
linear at low energies and shows deviations (sub- or super-sonic) at high 
energies (cf.~also \cite{Dumb1,Dumb2,Dumb3}).
Although there certainly exists a preferred frame in such systems, one might 
(formally) perform the same steps as described in the previous Section and
parameterize the non-linear dispersion relation $E(p)$ with a (non-unique) 
function $F$ as in Eq.~(\ref{function}) by ${\mathfrak E}^2=c^2{\mathfrak p}^2$
(but here $c$ denotes the speed of sound).
In this (somewhat artificial) way the usual (linear) Lorentz transformation 
$\cal L$ in the $({\mathfrak E},{\mathfrak p})$-space can be used to define 
transformations from one frame to another.
In order to Lorentz transform the field $\phi(t,x)$ (e.g., a wave-packet),
one first does a Fourier transform ${\cal F}$ assuming 
$ip\leftrightarrow\hbar\partial/\partial x$ etc., then applies the non-linear 
Lorentz transformation and finally transforms back:
\bea
\label{field}
\phi(t,x) &\to& \widetilde\phi(E,p)={\cal F}\phi
\,,
\nn
\widetilde\phi(E,p) &\to& \widetilde\phi(E',p')
\,,
\nn
\phi'(t,x) &=& {\cal F}^\dagger\widetilde\phi(E',p')
\,.
\ea
Since the function $F$ and its inverse $F^{-1}$ as well as the dispersion 
relation $E(p)$ are non-polynomial in general, the above procedure is clearly
non-local in position space $(t,x)$, see also Sections~{IV} and { V} 
below.

\section{Energy of composite systems}
%
As mentioned in the Introduction, one of the main motivations for deforming 
the (usual) Lorentz boosts is to require that not only the speed of light
(at low energies) but also the Planck scale is invariant\footnote{
In this sense the idea of DSR can be compared to the transition from the 
Galilei transformations (no invariant velocity) to special relativity, 
cf.~\cite{Amelino-DSR1,Amelino-DSR2}.} under the modified transformations 
(hence the notion ``doubly special relativity'').
Since the usual linear Lorentz boosts ${\cal L}$ do not possess any fixed 
points (in $E,p$) except zero and infinity and Eq.~(\ref{low-energy})  
connects $E=p=0$ to ${\mathfrak E}={\mathfrak p}=0$ the Planck scale must be 
mapped by the function $F^{-1}$ to infinity in order to be invariant.
As we shall see below, this property has rather dramatic consequences for 
composite systems.

Demanding that energy-momentum conservation in one frame has to be equivalent 
to energy-momentum conservation in all frames implies the following non-linear 
composition law\footnote{
The main argument is basically the same as for the usual linear Lorentz
transformations where (under certain assumptions, such as commutativity) 
only a linear composition law is invariant, see, e.g., \cite{Visser}.} 
\bea
\label{compose}
\left(\begin{array}{c}  E_{\rm total} \\ p_{\rm total} \end{array}\right)
=
F_{(N)}\left[\sum\limits_{i=1}^N
F^{-1}\left(\begin{array}{c}  E_i \\ p_i \end{array}\right)
\right]
\,,
\ea
i.e., one has to add the ${\mathfrak E}_i$ and ${\mathfrak p}_i$, 
see, e.g., \cite{Smolin-PRD,Visser}.
The subscript $(N)$ indicates that the function $F_{(N)}$ could be modified,
i.e., differ from $F$.

Let us first discuss the implications of using the same \cite{Visser} function 
as in the one-particle case $F_{(N)}=F$:
Since $F$ maps infinity to the Planck scale, the total energy can never exceed
the Planck energy -- which a weird result and raises serious questions 
concerning the physical significance of such an energy concept, see, e.g., 
\cite{Smolin-PRD}.
Moreover, a Galilei-type argument points out another contradiction if the 
velocity of propagation depends on the energy (see also the next Section):
If two or more particles have equal energies $E$ and hence velocities $v$ 
then the speed of the composite system obviously should be the same.
However, according to Eq.~(\ref{compose}) with $F_{(N)}=F$ the total energy 
$E_{\rm total}$ of the composite system is closer to the limiting Planck 
energy and hence the derived velocity would be different!

Alternatively, it has been suggested \cite{Smolin-PRD} that one constructs 
$F_{(N)}$ via replacing $M_{\rm Planck}$ by $NM_{\rm Planck}$ in the 
explicit expression for $F$.
In that case one has to know how many (elementary) particles the system is 
composed of -- which seems to be rather artificial, see, e.g., \cite{Visser}.
For example -- in addition to the arguments presented in \cite{Visser} -- 
if only one particle has the Planck energy (or very nearly so), 
adding one photon with an arbitrarily small energy to the system increases 
the total energy by $M_{\rm Planck}c^2$ -- which is also an odd feature.

In view of the above considerations, one might question the physical 
significance of the quantities $E$ and $p$ in comparison with 
${\mathfrak E}$ and ${\mathfrak p}$.
The laws of energy-momentum conservation assume a much simpler form in terms 
of ${\mathfrak E}$ and ${\mathfrak p}$ \cite{UHECR}.
Apparently the only justification for considering $E$ and $p$ instead of
${\mathfrak E}$ and ${\mathfrak p}$ could be that the former quantities, 
$E$ and $p$, are related to the space-time behavior and determine the 
(energy-dependent) velocity of propagation, etc., whereas the latter, 
${\mathfrak E}$ and ${\mathfrak p}$, are not.

\section{Energy-dependent speed of light}
%
Since the dispersion relation $E(p)$ must assume the same form in all frames,
it can be derived from the usual invariant ${\mathfrak E}^2=c^2{\mathfrak p}^2$
(remember $m=0$) of the linear Lorentz transformations ${\cal L}$.
Owing to the non-linear function $F$ the dispersion relation can 
involve a rather complicated dependence $E(p)$ with a possibly changing 
speed of light (cf.~the VSL cosmologies).
As indicated above, a varying velocity of propagation seems to be the only
possible way for the quantities $E$ and $p$ to acquire more physical 
significance than ${\mathfrak E}$ and ${\mathfrak p}$.

Obviously, the particle picture -- which the whole approach is based on -- 
and the concept of a velocity of propagation derived therefrom do only make 
sense if we are able to localize the particle under consideration with a 
(space/time) uncertainty much smaller than the length of the particle's 
world-line.
For example, we may derive the velocity of a Planckian particle by determining 
its position within a few Planck lengths and following its propagation over 
a macroscopic time duration and distance.
Here macroscopic means much larger than the Planck length/time (we want to 
retain the usual space-time translation symmetry and the concept of internal 
motion).

As motivated in the previous Section, the inverse function $F^{-1}$ 
diverges at the Planck scale and hence cannot be written as a polynomial 
(polynomials are regular everywhere). 
In general one would expect $E(p)$ to be singular at the Planck scale too
-- also displaying a non-polynomial behavior -- and therefore non-local 
effects to arise.
At a first glance, one might argue that these non-localities occur in the 
Planck regime only and are therefore not problematic.
However, as we shall demonstrate now, these non-local effects arise on a
{\em macroscopic} scale -- provided that the particles under consideration 
can travel a distance much larger than the Planck length 
(see the arguments above).

Let us consider the two limiting cases -- for higher and higher energies,
the speed of light goes to zero (sub-luminal dispersion) or to infinity 
(super-luminal).
In the first case, the particle basically stops moving and just sits there.
Now, if we can localize this highly Planckian particle within a few 
Planck lengths for a finite time duration 
(i.e., much longer than the Planck time) 
this clearly singles out a preferred frame, since we are supposed to know 
how Lorentz boosts act on macroscopic (i.e., sub-Planckian) scales!

In order to further study this apparent contradiction, 
let us consider a concrete example. 
Here one encounters a problem since, as mentioned in Section~{II},
the velocity of propagation is not uniquely determined.
In the following we assume that the speed of the particle is given by the 
group velocity $dE/dp$ (cf.~Section~{II} and \cite{c=dE/dp}) and 
choose a dispersion relation which is linear in some interval $p\in[p_1,p_2]$ 
-- though not with the usual proportionality factor $c$ -- say,
\bea
\label{disp-lin}
F\left(\begin{array}{c} E \\ p_1 \leq p \leq p_1 \end{array}\right)
&=&
\left(\begin{array}{c} E/2 \\ p \end{array}\right)
\,.
\ea
If we assume a very small boost velocity $v \ll c$ (Galilei limit), the 
Lorentz transformation in Eq.~(\ref{non-lin}) acts as
\bea
\label{half}
\left(\begin{array}{c}  E \\ p \end{array}\right)
\to
\left(\begin{array}{c}  E' \\ p' \end{array}\right)
&=&
\left(\begin{array}{c}  E+vp/2 \\ p+2vE/c^2 \end{array}\right)
\nn
&=&
\left(\begin{array}{cc} 1 & v/2 \\ 2v/c^2 & 1 \end{array}\right)
\left(\begin{array}{c}  E \\ p \end{array}\right)
\,,
\ea
i.e., with $v$ and $c$ being replaced by $v/2$ and $c/2$.

Since the dispersion relation $E^2=c^2p^2/4$ arising from 
Eq.~(\ref{disp-lin}) is linear (between $p_1$ and $p_2$), 
the Lorentz transformation in position space $(t,x)$ ought to be
the same as in momentum space $(E,p)$ -- no matter whether we 
consider a particle with $p_1 \leq p \leq p_2$ or a wave-packet 
(cf.~Section~{II}) with support in the interval $[p_2,p_3]$.
In this way the presence of the Planckian particle with 
$p_1 \leq p \leq p_2$ traveling over a long distance enforces 
a Lorentz boost with $v/2$ and $c/2$ instead of $v$ and $c$ 
-- at macroscopic (sub-Planckian) scales!

Evidently, the same phenomenon occurs for a super-luminal 
(speed of light goes to infinity) dispersion relation.
Any Planckian particle with a sub- or super-luminal velocity of 
propagation either introduces a preferred frame or necessitates the 
modification of the Lorentz transformation on its travel time and distance
-- i.e., on macroscopic (sub-Planckian) scales -- which demonstrates the 
occurrence of large-scale non-locality \cite{gedanken}.

\section{Loss of coincidence}
%
The fact that the presence of a Planckian particle affects the Lorentz 
transformations has further bizarre consequences.
If we go to 3+1 dimensions, the position-space representation of the deformed
Lorentz transformation described in Eq.~(\ref{field}) of Section~{II} 
acts as
\bea
\label{3D}
\phi'(t,\f{r}) 
&=& 
\int dE'd^3p'dt'd^3r'\,\phi(t',\f{r}')\times
\nn
&&\times\,\frac{1}{(2\pi)^4}\,
e^{i[tE'-\fk{r}\cdot\fk{p}'-t'E(E',\fk{p}')+\fk{r}'\cdot\fk{p}(E',\fk{p}')]}
\nn
&=&
\int dt'd^3r'\,{\mathfrak G}(t,t',\f{r},\f{r}')\,\phi(t',\f{r}')
\,.
\ea
The non-linearity in $E(E',\f{p}')$ and $\f{p}(E',\f{p}')$ results in a very 
strange behavior under space-time translations.
For the sake of illustration, we again (as in the previous Section) 
consider a function $F$ which is linear both, for low momenta and in some 
interval $p\in[p_1,p_2]$ 
\bea
\label{disp-lin-3D}
F\left(\begin{array}{c} E \\ \f{p}  \end{array}\right)
&=&
\left(\begin{array}{c} 
\left\{\begin{array}{lll}
E & : & \f{p}^2 \ll p_1^2 \\ 
E/2 & : & p_1^2 > \f{p}^2 > p_2^2
\end{array}\right.
\\ \f{p} \end{array}\right)
\,.
\ea
Now let us follow the evolution of two wave-packets -- 
one $\phi_{\rm low}'(t,\f{r})$ is decomposed of sub-Planckian energies 
$\f{p}^2 \ll p_1^2$ and the other one $\phi_{\rm Planck}'(t,\f{r})$ contains
momenta in the interval $p\in[p_1,p_2]$ only.
In this situation, the transformation in Eq.~(\ref{3D}) can be calculated 
easily, and in the Galilei limit $v \ll c$, we obtain, cf.~Eq.~(\ref{half})
\bea
\phi_{\rm low}'(t,\f{r}) 
&=& 
\phi_{\rm low}(t'+\f{r}'\cdot\f{v}/c^2,\f{r}'+\f{v}t')
\,,
\nn
\phi_{\rm Planck}'(t,\f{r}) 
&=& 
\phi_{\rm Planck}(t'+2\f{r}'\cdot\f{v}/c^2,\f{r}'+\f{v}t'/2)
\,.
\ea
Note that the relativistic corrections to the time coordinates in the first 
arguments on the right-hand side are different due to the non-linearity.
Consequently, if we change the origin of our spatial coordinate system,
we introduce a relative time shift 
\bea
\f{r}'\to\f{r}'+\f{a}\;\leadsto\;\Delta t=\f{a}\cdot\f{v}/c^2
\,,
\ea
between the two wave-packets.
Ergo, if the velocities of the two wave-packets and the boost direction
$\f{v}$ are linearly independent and the two wave-packets hit each other 
(i.e., coincide within their width at some space-time region) in one 
coordinate system, they may miss each other
(one wave-packet comes too late) in another coordinate representation!

Of course, this breaking of translational (i.e., Poincar\'e) invariance 
-- again on large scales -- 
has been demonstrated using the special field-theoretic representation 
described in Section~{II}; and one could argue that the above effect 
is an artifact of the special construction in Section~{II} and that 
in a different representation, this problem can be avoided. 
However, in order to prove this assertion, one has to provide another 
explicit field-theoretic example and to study its consequences.
It seems that one faces similar difficulties -- breaking of translational 
invariance $x_\mu\,\to\,x_\mu+a_\mu$ (see, e.g., \cite{Translation}) 
and deviations from the usual behavior on large scales 
$x_\mu \gg L_{\rm Planck}$ 
-- when introducing non-commuting coordinates via
\bea
[x^\mu,x^\nu]=\Lambda^{\mu\nu\rho}x_\rho
\,,
\ea
(instead of $[x^\mu,x^\nu]=\zeta\,g^{\mu\nu}$, for instance) as it is done 
for example in \cite{Non-commu1,Non-commu2} in relation to DSR theories.

We would also like to stress that the counter-argument presented in the 
previous Section is independent of any field-theoretic representation 
(and would therefore not go away).

\section{Summary}
%
Apart from the weird properties of composite systems discussed in 
Section~{III}, the theory of ``doubly special relativity'' goes along with 
violations of locality and separability if the speed of light depends on the 
energy, since the presence of a single Planckian particle can modify the 
action of the Lorentz transformation at macroscopic scales\footnote{
Note, however, that this result does not prove that DSR is conceptually 
inconsistent or in conflict with experiments or observations since we have 
not observed Planckian particles (at least not knowingly).} 
(i.e., much larger than the Planck length, see Sections~{IV} and { V}).
On the other hand, if the speed of light does not depend on the energy 
(e.g., the dispersion relation is $E^2=c^2p^2$), then there is no discernible 
reason to assign $E$ more physical significance than ${\mathfrak E}$, see
Section~{III}.
Although one should bear in mind that the whole approach purely relies on the 
particle picture (not a field theory), one would expect that the energy -- 
defined as the generator of the time-translation symmetry -- is an additive
quantity for independent systems (which brings us back to the question of 
locality and separability).

In search of alternatives one could imagine that -- even though the
(still to be found) underlying theory (including quantum gravity) might not
possess a preferred frame -- the physical state of the system describing the
actual gravitational field, etc., indeed does introduce an effectively
preferred frame with respect to the interaction with Planck-scale photons,
for example, that propagate within the gravitational field.

\section*{Acknowledgments}
%
R.~S.~acknowledges valuable conversations with G.~Volovik during a visit 
at the Low Temperature Laboratory in Finland, which was supported by the 
program EU-IHP ULTI 3. 
The authors also acknowledge support from the ESF-COSLAB program and 
thank the sponsors of Black Holes IV in Honey Harbor, Ontario.
This work was supported by 
the Alexander von Humboldt foundation, 
the Canadian Institute for Advanced Research, and 
the Natural Science and Engineering Research Council of Canada.
The authors thank G.~Amelino-Camelia, D.~Grumiller, R.~Lehnert, 
J.~Kowalski-Glikman, and L.~Smolin for discussions and comments.

\section*{Note added}
%
After finishing the work on our manuscript, we found that several other authors
(based on different approaches and assumptions) have also pointed out strange
consequences of DSR and/or concluded that DSR is either inconsistent with our 
present understanding of physics or trivial 
(i.e., indistinguishable from ordinary special relativity), 
see \cite{Lukierski,Rembielinski,Ahluwalia,Grumiller}.

\addcontentsline{toc}{section}{References}


\begin{thebibliography}{99}
\bibitem{UHECR}
At least for $F_{(N)}=F$ or for $N_{\rm initial}=N_{\rm final}$.
%
In these cases Eq.~(\ref{compose}) implies 
$\sum_i{\mathfrak E}_i^{\rm initial}=\sum_i{\mathfrak E}_i^{\rm final}$
as well as 
$\sum_i{\mathfrak p}_i^{\rm initial}=\sum_i{\mathfrak p}_i^{\rm final}$.
%
In view of this observation, it is hard to see how DSR can solve the threshold 
anomalies -- such as the UHECR problem -- which are basically based on energy 
conservation arguments, because one should be able to choose initial and final
configurations with the same number of particles; and at low energies, we have
${\mathfrak E}_i=E_i$.
%
(Unless, of course, one argues that the cross section changes significantly
at high energies, for example due to an energy-dependent speed of light.)
\end{thebibliography}
\end{document}